\documentclass[a4paper,fleqn,usenatbib]{mnras}

\usepackage{newtxtext,newtxmath}

\usepackage[T1]{fontenc}
\usepackage{ae,aecompl}

\usepackage{graphicx}	
\usepackage{amsmath}	
\usepackage{amssymb}	

\usepackage{hyperref}

\DeclareMathAlphabet{\mathsc}{OT1}{cmr}{m}{sc}
\def\testbx{bx}%
\DeclareRobustCommand{\ion}[2]{%
\relax\ifmmode
\ifx\testbx\f@series
{\mathbf{#1\,\mathsc{#2}}}\else
{\mathrm{#1\,\mathsc{#2}}}\fi
\else\textup{#1\,{\mdseries\textsc{#2}}}%
\fi}

\newcommand{\Heii}{\ion{He}{ii\ }}

\defcitealias{Hachisu1999}{HKNU99}

\title[Rates and delay times of SNe Ia]{Rates and delay times of Type I\lowercase{a} supernovae in the helium-enriched main-sequence donor scenario}

\author[Z.-W. Liu et al.]{
Zheng-Wei Liu$^{1,2,3,4}$\thanks{E-mail: zwliu@ynao.ac.cn} and Richard J. Stancliffe$^{4}$
\\
$^{1}$Yunnan Observatories, Chinese Academy of Sciences (CAS), Kunming 650216, P.R. China\\
$^{2}$Key Laboratory for the Structure and Evolution of Celestial Objects, CAS, Kunming 650216, P.R. China\\
$^{3}$Center for Astronomical Mega-Science, CAS, Beijing, 100012, P. R. China\\
$^{4}$Argelander-Institut f\"ur Astronomie, Auf dem H\"ugel 71, D-53121, Bonn, Germany\\
}

\pubyear{2017}

\begin{document}
\label{firstpage}
\pagerange{\pageref{firstpage}--\pageref{lastpage}}
\maketitle

\begin{abstract}

The nature of the progenitors of Type Ia supernovae (SNe Ia) remains a mystery. Comparing theoretical rates and delay-time distributions of SNe Ia with those inferred observationally can constrain their progenitor models. In this work, taking thermohaline mixing into account in the helium-enriched main-sequence (HEMS) donor scenario, we address rates and delay times of SNe Ia in this channel by combining the results of self-consistent binary evolution calculations with population synthesis models. We find that the Galactic SN Ia rate from the HEMS donor scenario is around $0.6$--$1.2\times10^{-3}\,\mathrm{yr^{-1}}$, which is about 30\% of the observed rate. Delay times of SNe Ia in this scenario cover a wide range of $0.1$--$1.0\,\mathrm{Gyr}$. We also present the pre-explosion properties of companion stars in the HEMS donor scenario, which will be helpful for placing constraints on SN Ia progenitors through analyzing their pre-explosion images.

\end{abstract}




\section{Introduction} \label{sec:introduction}

Type Ia supernovae (SNe Ia) play a fundamental role in astrophysics. They are important cosmological probes that led to the discovery of the accelerating expansion of the Universe \citep{Ries98, Perl99}. Although there is a general consensus that SNe~Ia arise from thermonuclear explosions of white dwarfs (WDs) in binary systems \citep{Hoyl60}, their specific progenitor systems have not yet been identified \citep[][]{Hillebrandt2000, Wang2012, Maoz2014}. The two leading scenarios proposed as SN Ia progenitors are the single-degenerate (SD, e.g., \citealt{Whel73, Han2004}) and double-degenerate (DD) scenario (e.g., \citealt{Iben84, Webbink1984}). In the SD scenario, a carbon-oxygen (CO) WD accretes matter from a non-degenerate companion, potentially a main-sequence (MS), subgiant (SG), red giant (RG), or even a helium (He) star, to trigger a  deflagration-to-detonation (e.g., \citealt{Khok91, Roep07, Seit12}), pulsational reverse detonation (e.g., \citealt{Bravo2006, Bravo2009a}), or gravitationally-confined detonation explosion (e.g., \citealt{Jordan2008, Jord12b, Seitenzahl2016}) when the WD mass comes close to the Chandrasekhar mass. In the DD scenario, an SN Ia explosion occurs after two WDs merge, in which the explosion can be triggered by a detonation either in the accretion stream (e.g., \citealt{Guillochon2010, Dan2012}) or in a violent merger involving massive WD \citep{Pakm10, Pakm11b, Pakm12b}. A He-accreting CO WD may also trigger a thermonuclear explosion before its mass reaches the Chandrasekhar-mass due to a detonation in the accreted He shell if the companion star is a He star or a He WD (e.g., \citealt{Woosley1986, Woosley2011, Bild07, Fink2010, Shen10, Shen14, Sim10, Krom10}). Alternative progenitor models for SNe Ia have also been suggested, including: the core-degenerate model which involves the merger of a WD with the core of an asymptotic giant branch star (e.g., \citealt{Livio2003, Kashi2011, Ilkov2012, Soker2013, Aznar2015}) and WD-WD head-on collisions either in dense stellar systems or in triple systems (e.g., \citealt{Benz1989, Raskin2009, Rosswog2009, Thompson2011, Kushnir2013}).


Certain recent observations may suggest that some SNe Ia might occur via the SD channel. For instance, it has been found that predicted early-time emissions due to the interaction of SN ejecta with a companion star \citep{Kasen2010, Liu2015c} can provide an explanation for the observed early light curves of SN~2012cg, iPTF14atg and SN~2017cbv. This seems to suggest that these three SNe were generated from SD binary systems \citep{Cao2015, Marion2016, Hosseinzadeh2017}. However, the SD origin of these three events has been questioned by other studies \citep{Kromer2016, Liu2016, Shappee2016, Levanon2017, Noebauer2017}. Because the CSM is generally expected to exist around SN Ia progenitor as the result of mass transfer from the companion star, as well as from WD winds, detecting some narrow absorption signatures of circumstellar material (CSM) in some SNe Ia has been suggested to be evidence that supports the SD scenario  (\citealt{Pata07, Ster11, Dild12, Silverman2013}, but see \citealt{Soker2013}). On the other hand, there is some evidence in favour of the DD scenario, e.g., the non-detection of pre-explosion companion stars in normal SNe Ia \citep{Li2011, Bloom2012, Kell14}, the lack of radio and X-ray emission around peak brightness \citep{Chom12,  Hore12, Marg14}, the absence of a surviving companion star in SN Ia remnants \citep{Kerz09, Scha12, Ruiz2017}, and no signatures of the swept-up H/He predicted by hydrodynamical models \citep{Marietta2000, Liu2012, Liu2013a, Liu2013c, Pan2012} have been detected in the nebular spectra of SNe Ia \citep{Leon07, Lund13, Maguire2016}.

Comparing the expected rates and delay-time distributions of SNe from binary population synthesis calculations with those inferred from the observations have been widely used to place constraints on progenitor systems of SNe Ia (\citealt{Yungelson1998, Nelemans2001, Han2004, Botticella2008, Mannucci2008, Ruiter2009, Maoz2010, Meng2010, Wang2010, Mennekens2010, Bours2013, Claeys2014, Graur2014, Liu15a, Shen2017}). In particular, the MS donor scenario, which is sometimes known as the ``WD+MS'' or the supersoft channel, has been comprehensively investigated by many studies. It has been found that the MS donor scenario is one of the most efficient SD scenarios for producing SNe Ia (e.g. \citealt{Hachisu1999, Han2004, Meng2010, Wang2010}). In this scenario, a WD accretes H-rich material from a MS or a slightly evolved SG companion star to grow in mass until the Chandrasekhar-mass limit is reached, triggering an SN Ia. The rate of SNe Ia originating from WD+MS systems in our Galaxy has been found to be around $0.2$--$1.8\times10^{-3}\,\mathrm{yr^{-1}}$, which can account for $\lesssim2/3$ of the observed rate (which is around $3\times10^{-3}\,\mathrm{yr^{-1}}$, \citealt{Cappellaro1997}). Also, SNe Ia in this scenario have a relatively long delay time of about $250\,\mathrm{Myr}$--$1\,\mathrm{Gyr}$.

Depending on the evolutionary state of the primary when the binary system undergoes the first RLOF event, there are three evolutionary channels for forming CO~WD+MS binary systems \citep{Wang2012}:\vspace{-\topsep}
\begin{itemize}
\item[(i)] $HG/RGB$ channel: the binary system starts the first episode of RLOF mass transfer as the primary fills its Roche lobe in the Hertzsprung gap (HG) or on the red-giant branch (RGB) phase. The binary system then experiences an unstable mass transfer and undergoes a common envelope (CE) event, forming a He star and a MS companion. This He star subsequently overfills its Roche lobe during core He burning to deposit He-rich material onto its MS companion. As a consequence, a binary system consisting of a CO~WD and a MS companion star is produced.
\item[(ii)] $EAGB$ channel: the system experiences dynamically unstable mass transfer as the primary fills its Roche lobe at the early asymptotic giant branch (EAGB) stage. As a result, the system enters the CE phase. After the CE is ejected, a binary system consisting of a He RG and a MS companion star is produced. The He RG subsequently fills its Roche lobe and starts stable mass-transfer, resulting in a CO WD+MS system. 
\item[(iii)] $TPAGB$ channel: the primordial primary fills its Roche lobe during the thermally-pulsing asymptotic giant branch (TPAGB) stage. The system then enters the CE phase. A CO~WD+MS system is directly formed after the binary undergoes the CE event.
\end{itemize}

\begin{figure*}
  \begin{center}
    {\includegraphics[width=0.88\textwidth, angle=360]{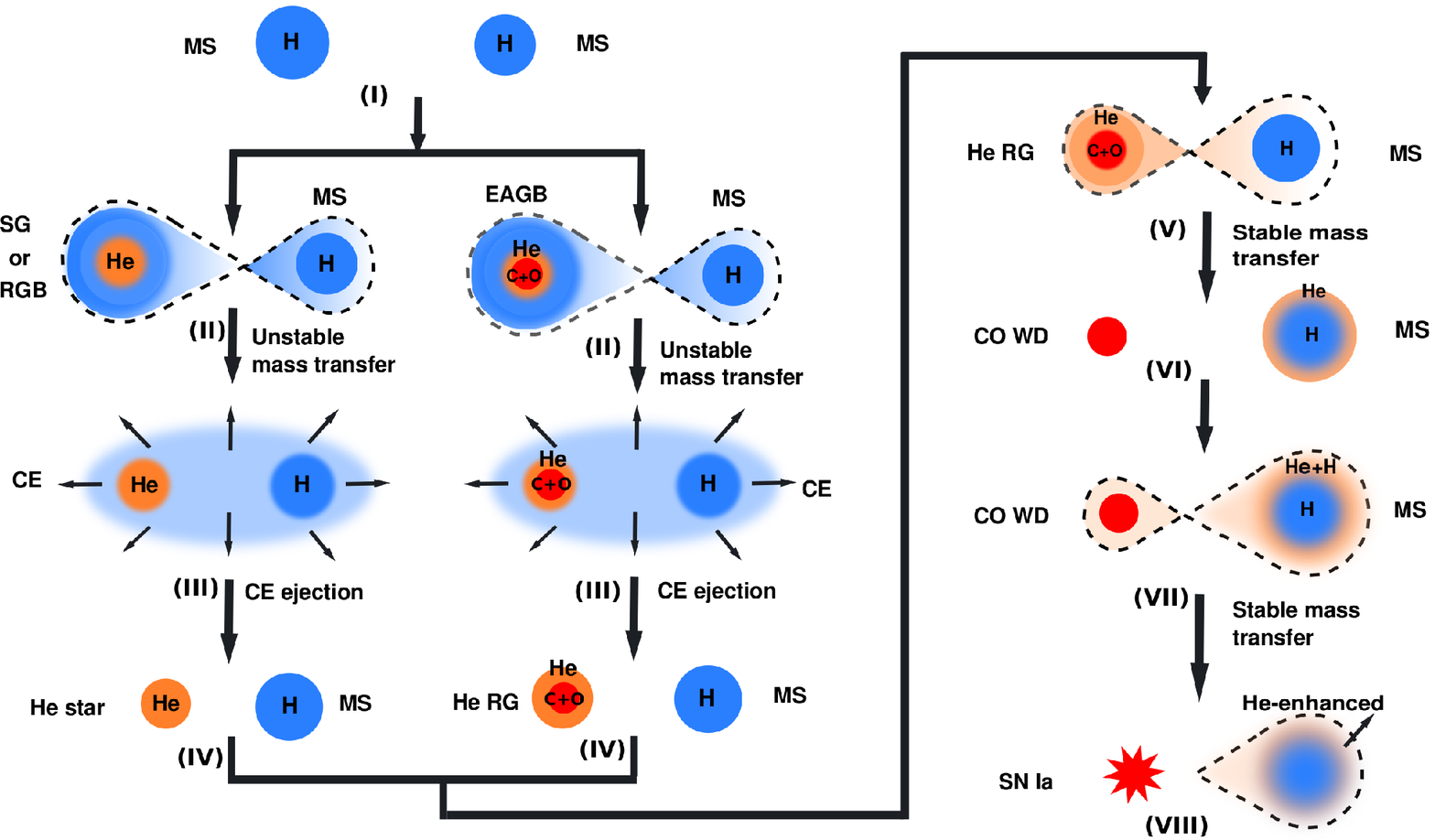}}
    \caption{Binary evolutionary path of the HEMS donor scenario of SNe Ia. Depending on the evolutionary state of the mass donor star when the system undergoes the first RLOF mass transfer, the outcome of a CE event can be a He~star+MS system or a He~RG+MS system (see also \citealt{Hachisu1999}).}
\label{Fig:1}
  \end{center}
\end{figure*}

A detailed schematic of the evolution of a binary system in the \textit{HG/RGB} channel and \textit{EAGB} channel is shown in Fig.~\ref{Fig:1}. As shown, the donor stars of the WD+MS systems in these two channels are expected to have He-enriched outer layers as a significant amount of pure He material has been transferred from a He star or He RG during the previous RLOF phase (stage~$\mathrm{V}$  of Fig.~\ref{Fig:1}, see also \citealt{Hachisu1999}). Rather than leave a layer of almost pure He on the surface of the MS companion, the material accreted from the He star or He RG will become mixed into the companion's interior by the action of thermohaline mixing. This process occurs when the mean molecular weight of the stellar gas increases towards the surface, in this case because the accreted He layer has a higher mean molecular weight than the material of the MS companion. A gas element, displaced downwards and compressed, will be hotter than its surroundings. It will therefore lose heat, increase in density and continue to sink. This results in mixing on a thermal timescale until the molecular weight difference has disappeared \citep{Kippenhahn1980, Stancliffe2007}.  The effects of thermohaline mixing on the structure and composition of stars have been widely analyzed either in low-mass binaries or in massive systems (e.g., \citealt{Wellstein2001, Stancliffe2007, Stancliffe2008, Stancliffe2010}), as thermohaline mixing is expect to naturally occur in binary systems when material that has undergone nuclear processing from the primary star is transferred to its less evolved secondary.

The He-rich material transferred to the surface of the MS secondary at stage~$\mathrm{V}$ (Fig.~\ref{Fig:1}) alters the surface composition of the secondary because thermohaline mixing is expected to naturally occur as this accreted material has a greater mean molecular weight than that of the MS secondary \citep{Kippenhahn1980}. Therefore, He-enrichment of the MS companion star and thermohaline mixing cannot be neglected in the $HG/RGB$ and $EAGB$ channel. However, these two effects have not generally been taken into account by current population synthesis studies for the CO~WD+MS scenario which treated the MS star as a normal solar-metallicity MS star. Here, we call the $HG/RGB$ and $EAGB$ channel the helium-enriched MS (HEMS) donor scenario of SNe Ia, i.e., the CO~WD+HEMS scenario \citep{Liu2017}.

\begin{figure}
  \begin{center}
    {\includegraphics[width=0.45\textwidth, angle=360]{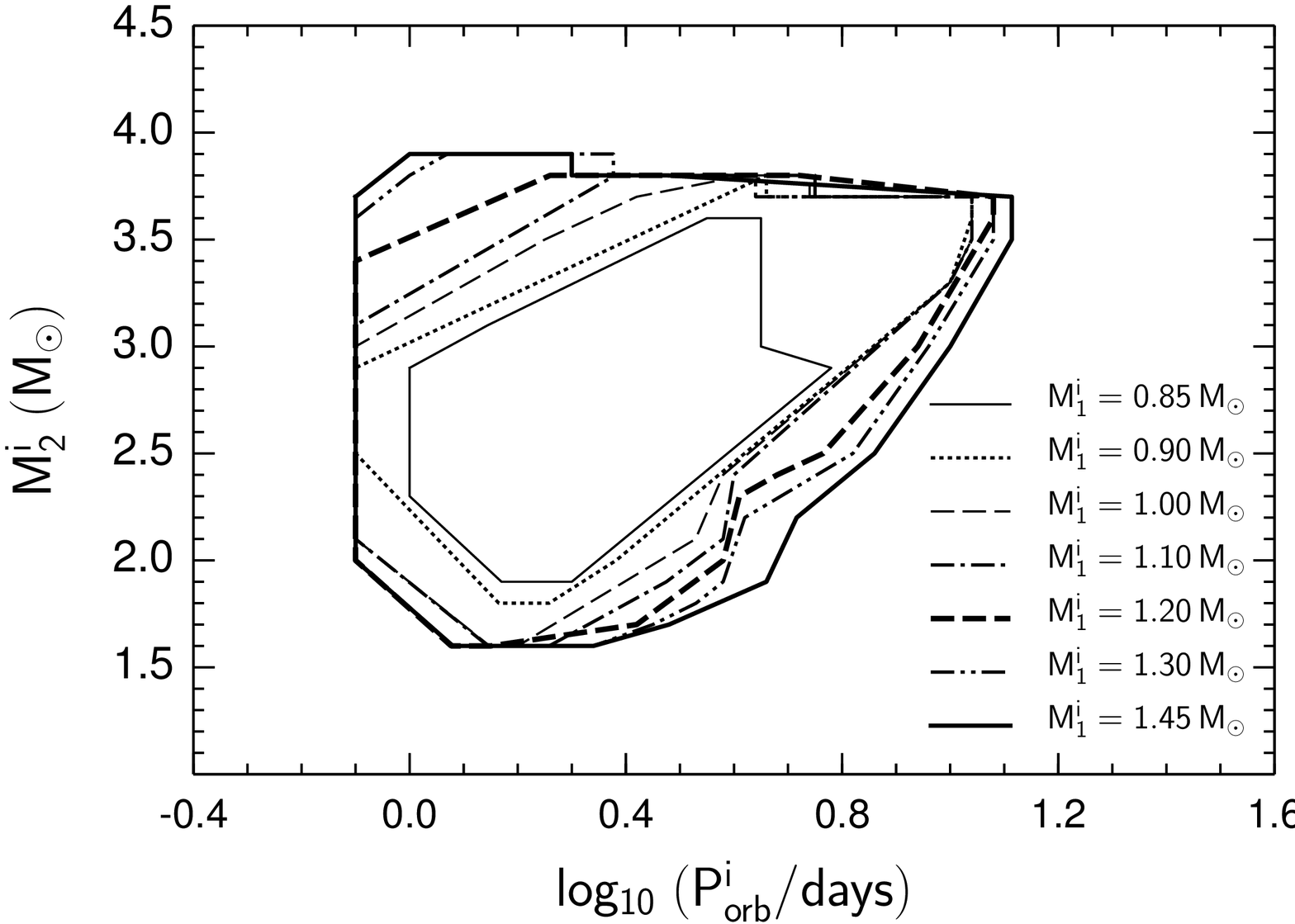}}
    \caption{Regions in the orbital period-secondary mass plane for He star+MS binary systems (stage $\mathrm{IV}$ of Fig.~\ref{Fig:1}) that successfully produce SNe Ia based on our consistent binary evolution calculations. Different curves correspond to regions for different He star masses.}
\label{Fig:contour}
  \end{center}
\end{figure}

We revise the rates and delay times of SNe Ia in the HEMS donor scenario in this work. We particularly concentrate on the $HG/RGB$ evolutionary channel because previous binary population synthesis studies have been found that this channel is the most significant route for producing SNe Ia among the three evolutionary channels mentioned above \citep{Meng2010, Wang2012}. With self-consistent binary evolution calculations, \citet{Liu2017} have found that the amount of hydrogen swept-up by the SN ejecta in the HEMS donor scenario is higher than observational limits on the hydrogen masses in progenitors of SNe Ia, which is in conflict with the fact that no strong signatures of swept-up hydrogen have been detected in nebular spectra of SNe Ia (e.g. \citealt{Leon07, Maguire2016}). If the HEMS donor scenario is a significant route to SNe Ia formation, then the non-detection of hydrogen in these systems is a serious problem. Therefore, determining the rate of SNe Ia in this scenario from an updated population synthesis calculation which includes the effect of thermohaline mixing is needed.

\section{Numerical Methods and Results}
\label{sec:method}

To predict the rate and delay times of SNe Ia in the HEMS donor scenario, we use the method adopted by \citet{Han2004}, in which the results of detailed binary evolution calculations are combined into population synthesis models. First, we perform self-consistent binary evolution calculations for a set of binary systems from stage~$\mathrm{IV}$ to $\mathrm{VIII}$ in Fig.~\ref{Fig:1} to obtain initial parameter contours at stage~$\mathrm{IV}$ which finally lead to a successful SN Ia, i.e., $M_{1}^{\,\mathrm{i}}$, $M_{2}^{\,\mathrm{i}}$ and $P_{\mathrm{orb}}^{\,\mathrm{i}}$. These initial parameter contours are then interpolated into population synthesis models to determine rates and delay times of SNe Ia: once a binary system in calculations with a rapid binary evolution code evolves to stage~$\mathrm{IV}$ and falls into the initial parameter contours ($M_{1}^{\,\mathrm{i}}$, $M_{2}^{\,\mathrm{i}}$ and $P_{\mathrm{orb}}^{\,\mathrm{i}}$) obtained from our binary evolution calculations, it is assumed to lead to a successful SN Ia. Details on how we set up basic parameters in the codes and on our most important assumptions of self-consistent binary evolution and population synthesis calculations are described in the next sections.

\subsection{Binary Evolution Calculation}
\label{sec:binary}

The Cambridge stellar evolution code {\sc STARS} \citep{Eggl71, Eggl72, Pols1995, Stancliffe2009} is used to trace the detailed binary evolution for the HEMS donor scenario. We start our calculation when a massive primary star evolves to a He star, i.e., at stage $\mathrm{IV}$ in Fig.~\ref{Fig:1}. Thermohaline mixing has been implemented into the code as described by \citet{Stancliffe2007}. The detailed structures of two components in the binary system are consistently solved in our calculation from stage $\mathrm{IV}$ to $\mathrm{VII}$. Once the He primary star evolves to become a WD and the He-enriched MS companion starts to fill its Roche lobe at stage $\mathrm{VII}$, instead of solving the detailed structure of the WD, we treat the WD as a point mass and follow the method of \citet{Hachisu1999} to calculate the mass growth rate of the WD, $\dot{M}_{\rm{WD}}$. We set up the mass accumulation efficiency for hydrogen shell burning $\eta_{\rm{H}}$ as follows:
 \begin{equation}
    \label{eq:1}
\footnotesize
\begin{array}{l}
\eta_{\rm{H}} = \\[0.5em]
\left\{ \begin{array}{ll}
\mathrm{CE\ event}\,, & \ \ \ \ \ \ \   (|\dot{M}_{\rm{2}}| > 10^{-4}\,M_{\sun}\,\mathrm{yr^{-1}}) \\[0.4em]
\dot{M}_{\rm{cr}}/|\dot{M}_{\rm{2}}|\,,  & \ \ \ \ \ \ \ (10^{-4}\,M_{\sun}\,\mathrm{yr^{-1}} > |\dot{M}_{\rm{2}}| > \dot{M}_{\rm{cr}})  \\[0.4em]
1\,, & \ \ \ \ \ \ \ (\dot{M}_{\rm{cr}} \geqslant |\dot{M}_{\rm{2}}| \geqslant 10^{-7}\,M_{\sun}\,\mathrm{yr^{-1}})\\[0.4em]
0\,, & \ \ \ \ \ \ \ (|\dot{M}_{\rm{2}}| < 10^{-7}\,M_{\sun}\,\mathrm{yr^{-1}})
\end{array} \right.
\end{array}
  \end{equation}
where $\dot{M}_{\mathrm{cr}} = 1.2\times10^{-6}(M_{\mathrm{WD}}/M_{\odot}-0.4)\,M_{\sun}\,\mathrm{yr^{-1}}$ is the critical accretion rate for stable hydrogen burning, $M_{\rm{WD}}$ is the mass of the accreting WD, $\dot{M}_{\rm{2}}$ is the mass transfer rate, and $X$ is the hydrogen mass fraction. The optically thick wind \citep{Hachisu1999} is assumed to blow off all unprocessed material if $|\dot{M}_{\rm{2}}| > \dot{M}_{\rm{cr}}$, and the lost material is assumed to take away the specific orbital angular momentum of the accreting WD.

\begin{figure}
  \begin{center}
    {\includegraphics[width=0.45\textwidth, angle=360]{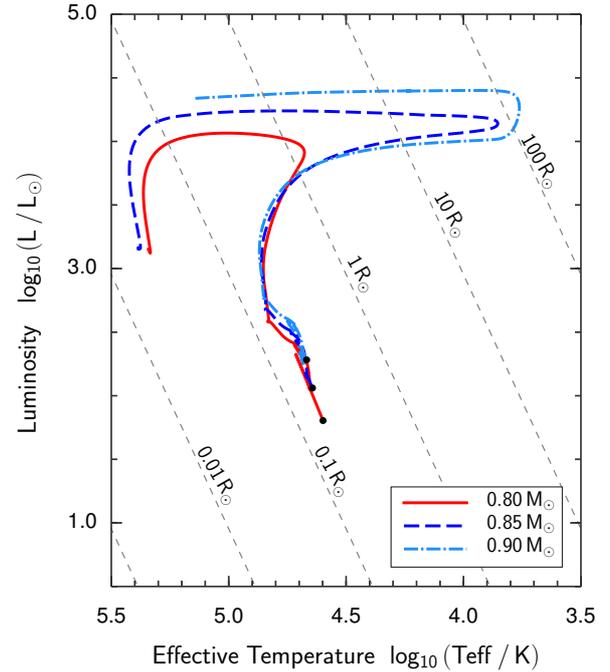}}
    \caption{Hertzsprung-Russell diagram showing the evolutionary tracks of the He stars in our detailed binary evolution calculations from stage IV to stage VI for three example models with an initial He star mass of $0.80\,M_{\sun}$, $0.85\,M_{\sun}$ and $0.90\,M_{\sun}$, respectively.}
\label{Fig:hr}
  \end{center}
\end{figure}

\begin{figure*}
  \begin{center}
    {\includegraphics[width=0.43\textwidth, angle=360]{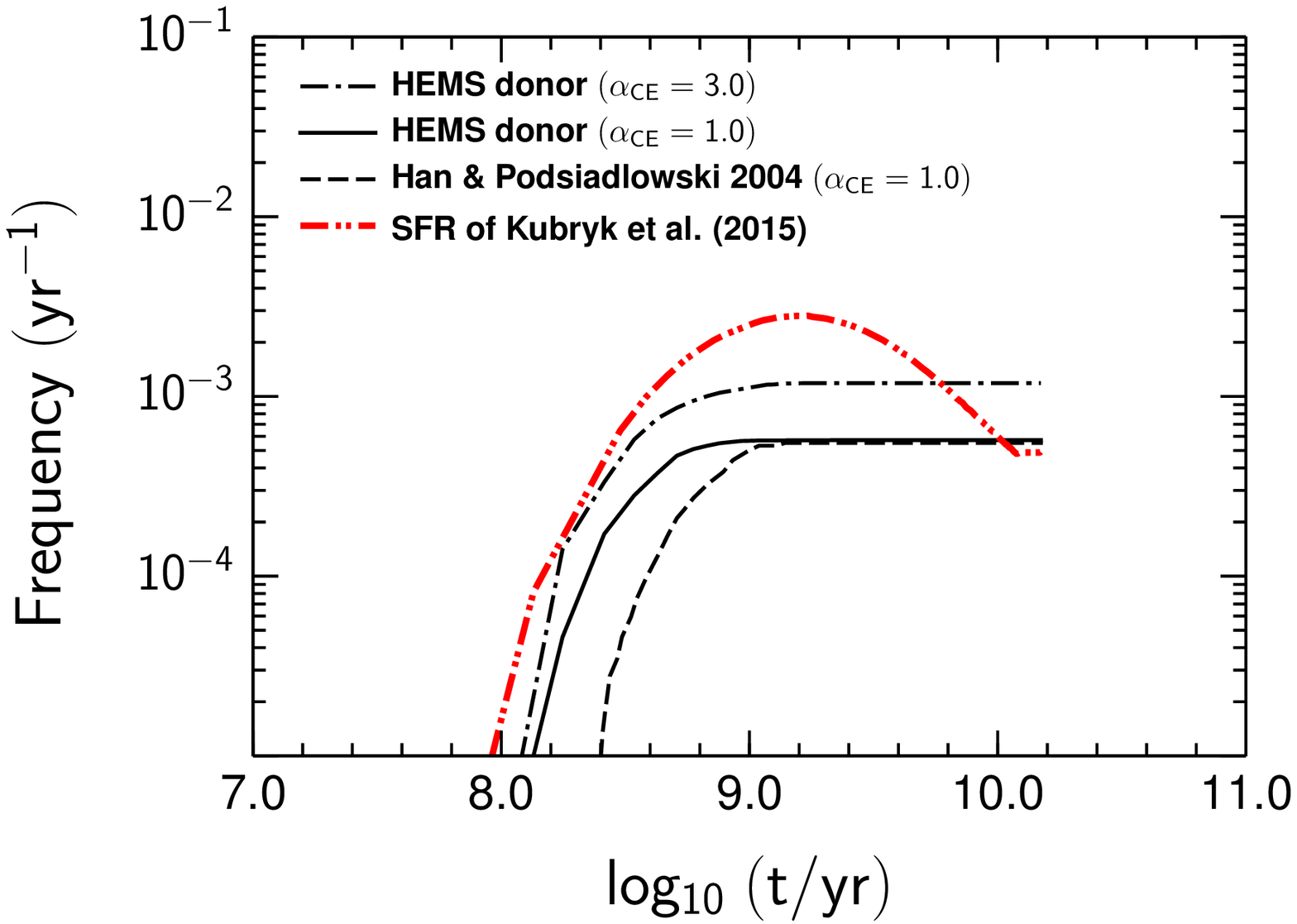}}
   \hspace{0.3in}
    {\includegraphics[width=0.43\textwidth, angle=360]{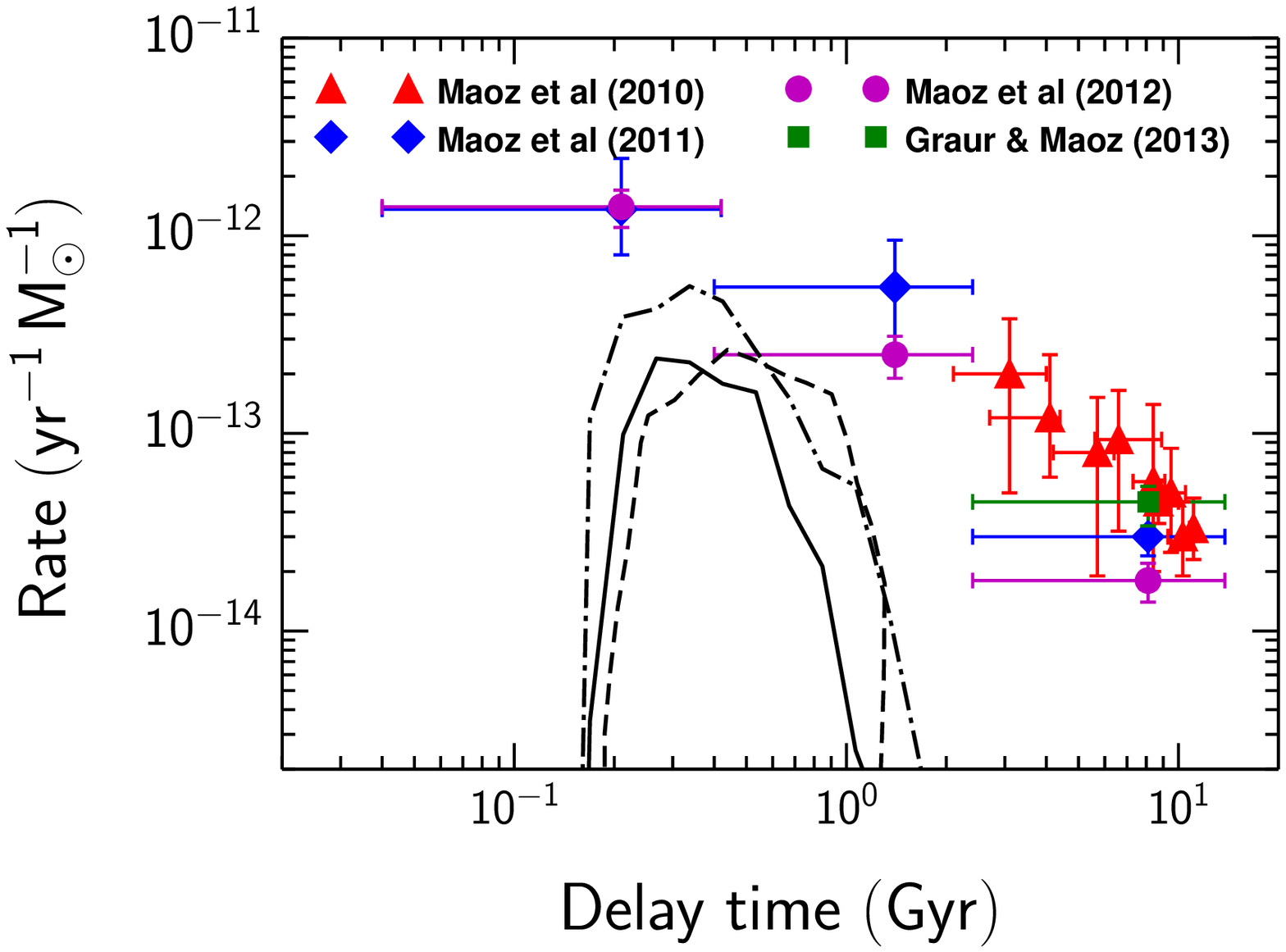}}
    \caption{Left-hand panel: Galactic rates of SNe Ia as a function of delay time in the HEMS donor scenario for a constant star formation rate ($Z$=$0.02$, SFR=$5\,M_{\sun}\,\mathrm{yr^{-1}}$). The red double-dotted curve gives the result when adopting the SFR of the Milky Way given by \citet[][see their Fig.~2]{Kubryk2015}, in which the total SFR of the Milky Way is decomposed into Bulge and Disk contributions (see also \citealt{Maoz2017}). Right-hand panel: the measured DTD of the HEMS donor scenario in our calculations. Here, we also compare our results to the observed DTD \citep{Maoz2010, Maoz2011, Maoz2012, Maoz2017, Graur2013}. For a comparison, the results of \citet{Han2004} are given by the dashed curves.}
\label{Fig:rate}
  \end{center}
\end{figure*}

\begin{figure*}
  \begin{center}
    {\includegraphics[width=0.43\textwidth, angle=360]{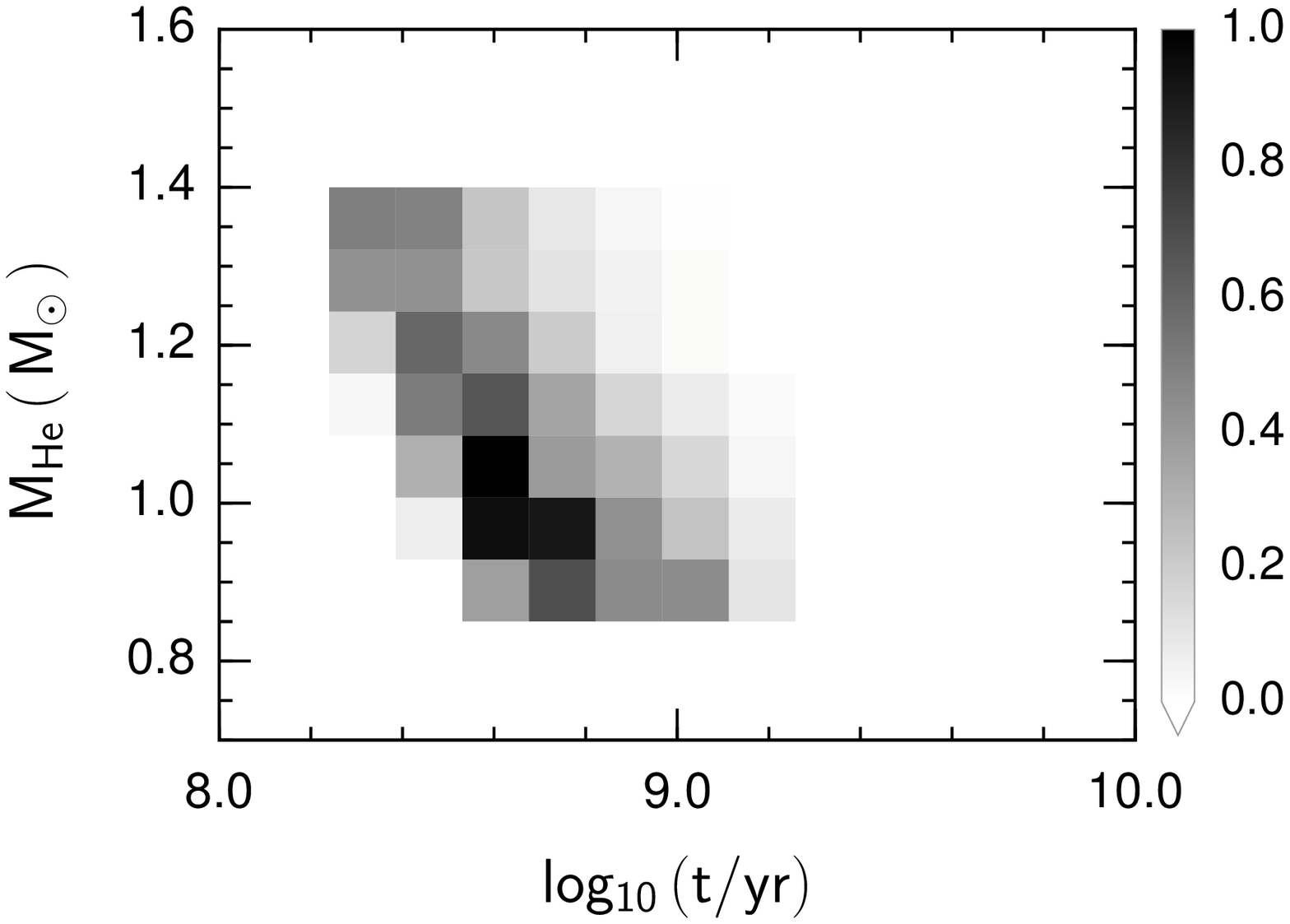}}
   \hspace{0.3in}
    {\includegraphics[width=0.43\textwidth, angle=360]{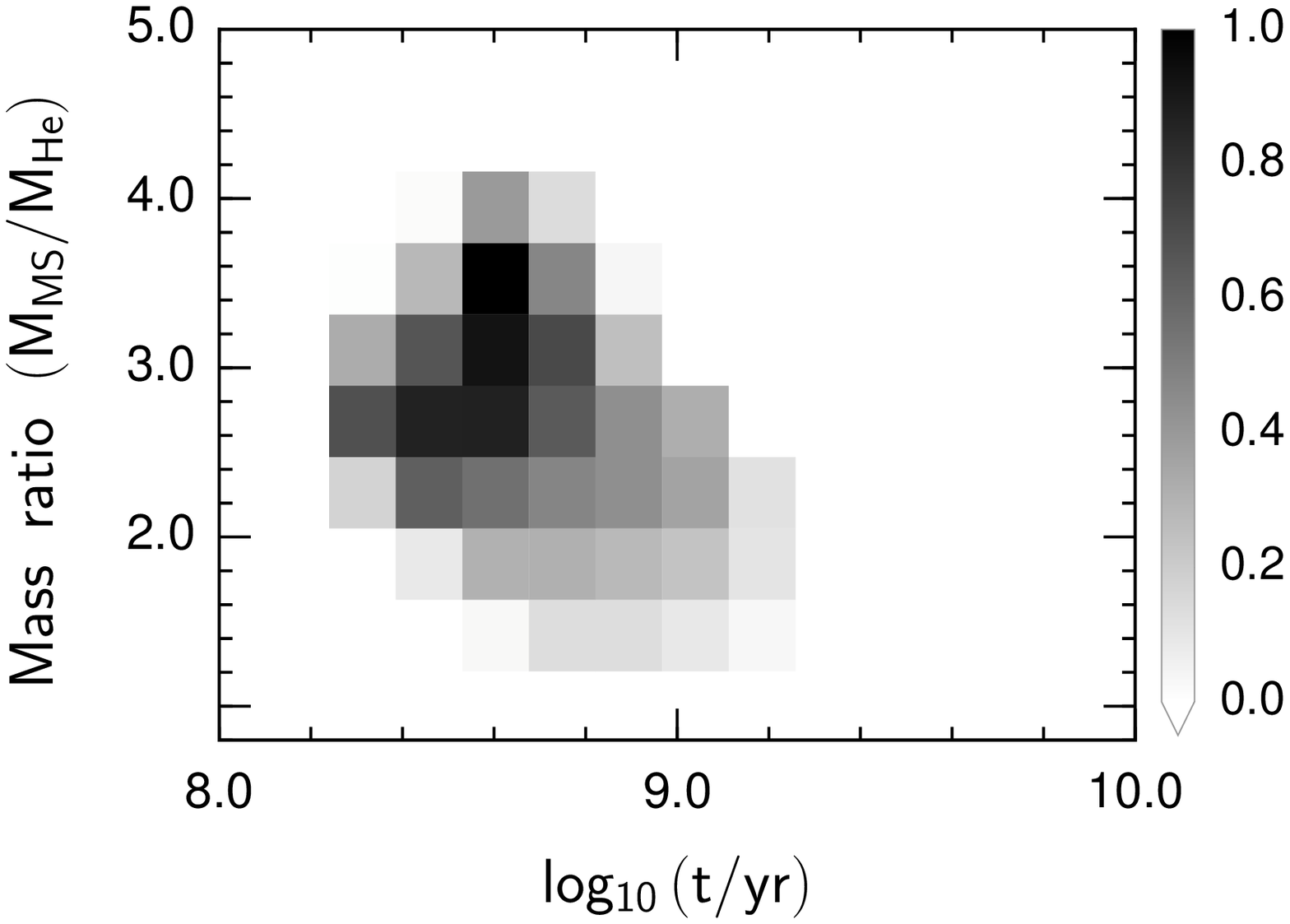}}
    \caption{Mass of a He star (left panel) and the mass ratio of He~star+MS binary systems (right panel) at stage~IV (Fig.~\ref{Fig:1}) as a function of delay times of SNe Ia in our population synthesis calculations with a CE ejection efficiency of $\alpha_{\mathrm{CE}}$=$3.0$. The results with $\alpha_{\mathrm{CE}}$=$1.0$ are similar.}
\label{Fig:initial}
  \end{center}
\end{figure*}

The mass accumulation efficiency for helium shell burning $\eta_{\rm{He}}$, is calculated based on \citet{Kato1999}: 
 \begin{equation}
    \label{eq:2}
\footnotesize
\begin{array}{l} 
\eta_{\mathrm{He}} = \\[0.5em]
\left\{\begin{array}{ll}
1,&(-5.9 \leqslant \mathrm{log}\,\dot{M}_{\mathrm{He}}\lesssim-5.0)\\[0.3em]
-0.175\,(\mathrm{log}\,\dot{M}_{\mathrm{He}}+5.35)^{2}+1.05,&(-7.8\leqslant\mathrm{log}\,\dot{M}_{\mathrm{He}}<-5.9)
\end{array}\right.
\end{array}
  \end{equation}

We assume the WD explodes as an SN Ia,  i.e., at stage $\mathrm{VIII}$ in Fig.~\ref{Fig:1}, when its  masses reaches the Chandrasekhar-mass limit (which we take as $1.4\,M_{\sun}$). Any rotation of the WD is not considered in our calculations.

\subsection{Outcomes of the binary evolution calculations}

Adopting the method described in Section~\ref{sec:binary}, we performed consistent binary evolution calculations for around 15000 binary systems consisting of a He star and a MS companion star. These initial binary systems have a primary star of $0.8$--$1.45\,M_{\sun}$ and a secondary of $1.2$--$4.2\,M_{\sun}$ with an orbital period of $0.6$--$16\,\mathrm{days}$. Fig.~\ref{Fig:contour} presents the initial parameter spaces of He star+MS binary systems in the orbital period-secondary mass plane ($\mathrm{log_{10}}\,P_{\mathrm{orb}}^{i}$--$M_{2}^{i}$, which corresponds to stage IV in Fig.~\ref{Fig:1}) which can successfully produce SNe Ia in our binary evolution calculations. At low periods, the contour (the left boundaries in Fig.~\ref{Fig:contour}) is constrained by requiring that the systems are not filling their Roche lobes at the beginning of the simulation. The systems outside the upper and right boundaries of contours mainly produce WD+HEMS systems which subsequently experience dynamically unstable mass transfer and eventually lead to common envelope objects. For the WD to be able to reach the Chandrasekhar limit, the donor star should be massive enough and the mass-transfer rate has to be sufficiently high. These constraints give the lower boundaries here. Our calculations cover a mass range of the initial He star from $0.8\,M_{\sun}$ to $1.45\,M_{\sun}$. Carbon ignition takes place in He stars of more than $1.45\,M_{\sun}$, preventing them forming CO WDs (see also \citealt{Hachisu1999}). Therefore, a value of $1.45\,M_{\sun}$ is set to be the upper-limit mass of a He star in our binary evolution calculations. On the other hand, a He star with a mass of $0.8\,M_{\sun}$ does not expand significantly during its lifetime  (Fig.~\ref{Fig:hr}). It is therefore unlikely to fill its roche lobe to transfer He-rich material to the companion star and thus produce a WD+HEMS binary system. This means that this system does not experience stage~V in Fig.~\ref{Fig:1}. As a result, no system in our grid with a $0.8\,M_{\sun}$ He star can successfully form a WD+HEMS binary system and lead to an SN Ia. Therefore, $0.85\,M_{\sun}$ is set to be the lower-limit mass of a He star in our calculations.

\begin{figure*}
  \begin{center}
    {\includegraphics[width=0.3\textwidth, angle=360]{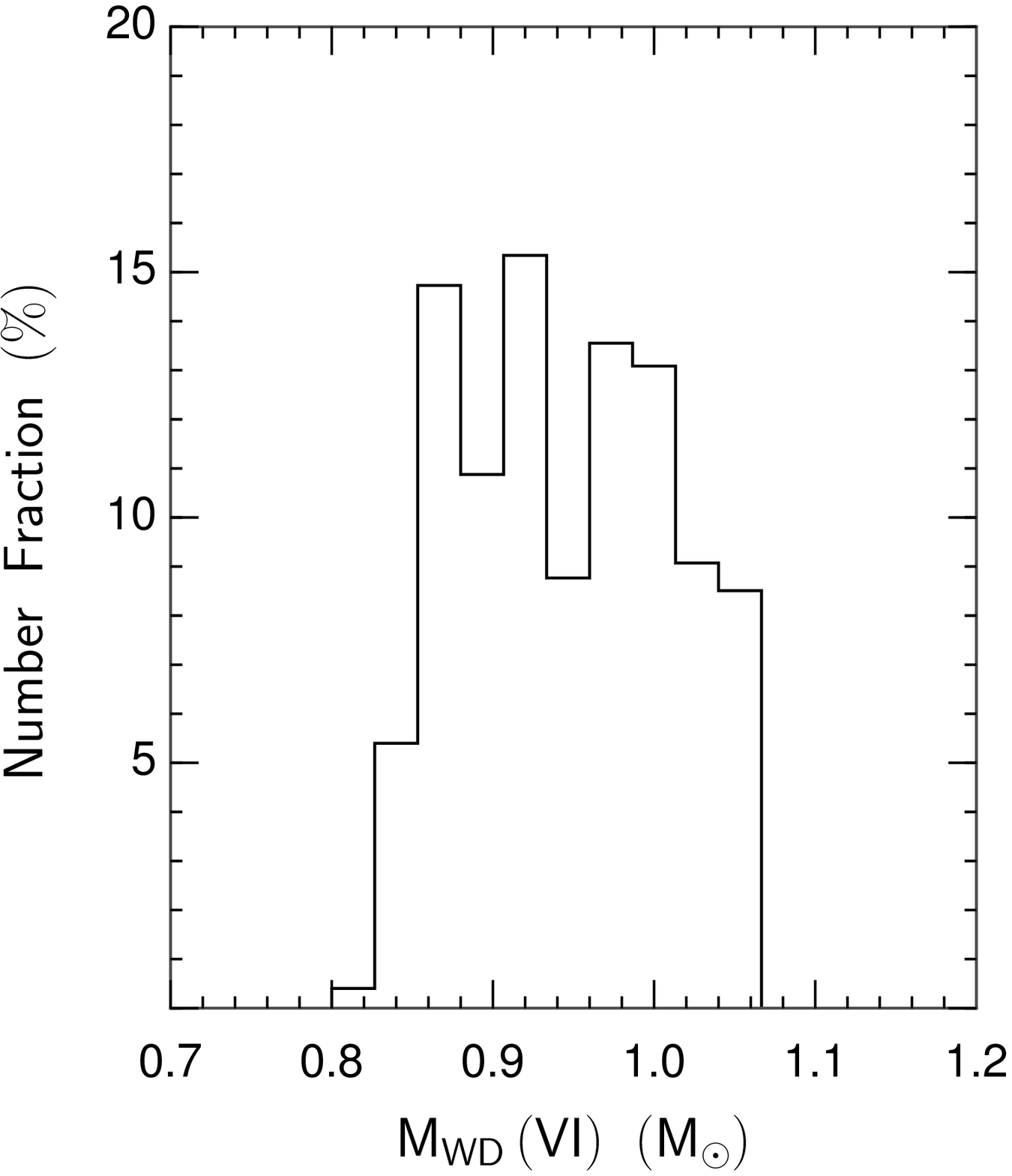}}
   \hspace{0.1in}
    {\includegraphics[width=0.3\textwidth, angle=360]{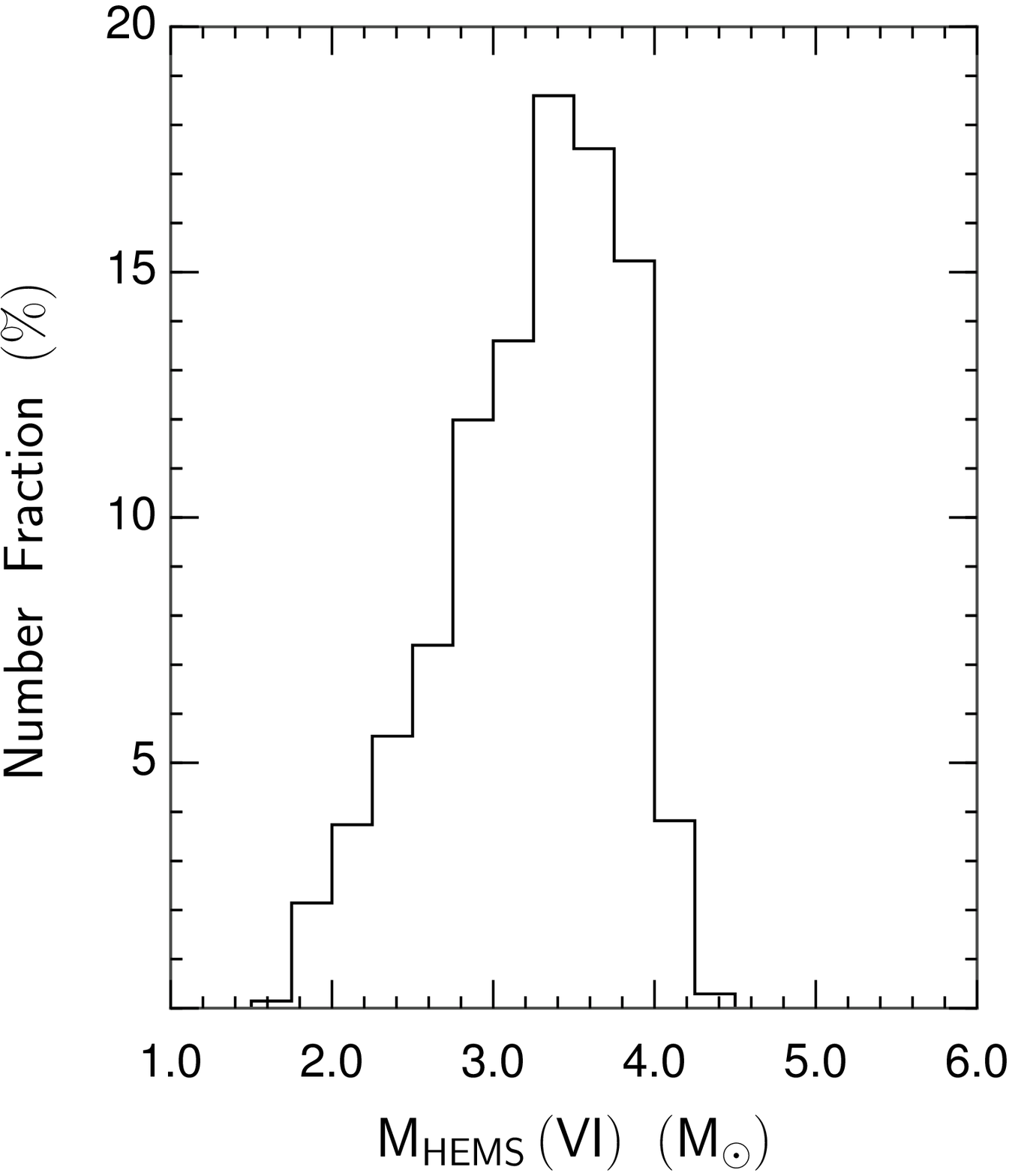}}
   \hspace{0.1in}
    {\includegraphics[width=0.3\textwidth, angle=360]{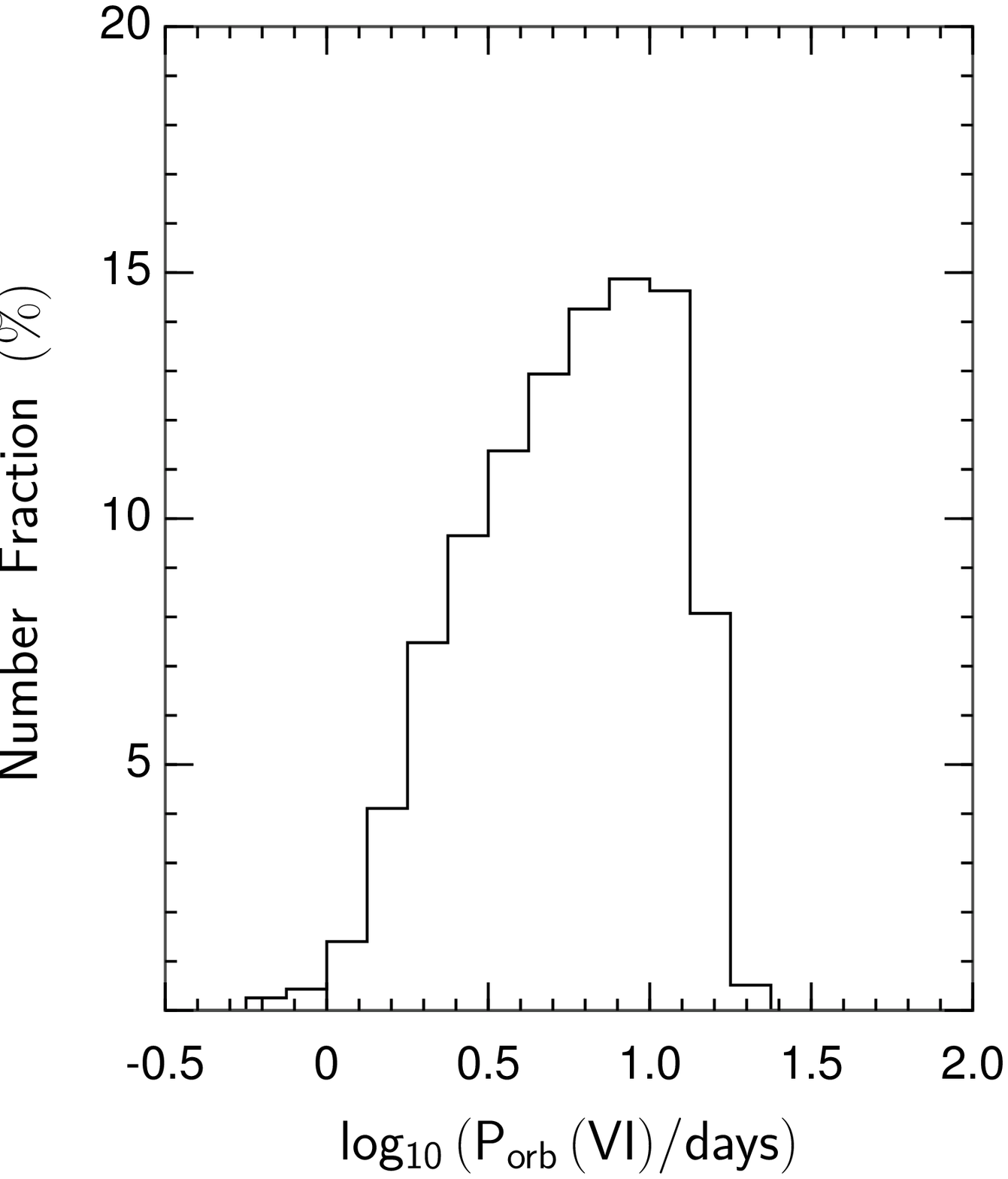}}
    \caption{Distributions of binary parameters of CO WD+HEMS systems (stage~VI in Fig.~\ref{Fig:1}) which successfully produce SNe Ia based on our consistent binary evolution calculations with the {\sc STARS} code (Section~\ref{sec:binary}). $M_{\mathrm{WD}}$(VI),  $M_{\mathrm{HEMS}}$(VI) and $P_{\mathrm{orb}}$(VI)  are the mass of the CO WD, the mass of the HEMS companion star and the orbital period of the binary system, respectively.}
\label{Fig:binary}
  \end{center}
\end{figure*}

\subsection{Population Synthesis Calculation}
\label{sec:bps}

In this work, rates and delay times of SNe Ia in the HEMS donor scenario are predicted by performing population synthesis calculations with the rapid binary evolution code developed by \citet{Hurley2000, Hurley2002}\footnote{This rapid binary evolution code is freely available on \url{http://astronomy.swin.edu.au/~jhurley/}.}. Basic parameters and assumptions in our population synthesis calculations are set to be consistent with those of \citet{Han2004}, which are described in detail below.  

\subsubsection{Common-Envelope Evolution}
\label{sec:CE}

The details of mass transfer at stage~$\mathrm{II}$ in Fig.~\ref{Fig:1} are determined by the mass ratio of the binary system, i.e., the ratio of the mass of the donor star to the mass of the accretor ($M_{2}/M_{1}$). The binary system undergoes a dynamically unstable mass transfer and enters a CE event if the mass ratio of the system is larger than the critical mass ratio, $q_{\,\mathrm{crit}}$ \citep{Paczynski1976}. Depending on the evolutionary state of the donor star at the onset of RLOF, the treatment on the value of $q_{\,\mathrm{crit}}$ is different, which has been discussed in detail by many studies (e.g., \citealt{Hjellming1987, Webbink1988, Hurley2002, Han2002}). In this work we use $q_{\,\mathrm{crit}}=4.0$ when the donor star is on the MS stage or in the HG \citep{Han2000, Hurley2002}. If the donor star is a naked He giant, we set $q_{\,\mathrm{crit}}$ to be 0.784. If the donor star fills its Roche lobe on the first giant branch (FGB) or AGB, $q_{\,\mathrm{crit}}$ is given as follows:
\begin{equation}
q_{\,\mathrm{crit}} = \left. \left [\,1.67 - x + 2\,\left (\left. M_{\mathrm{1}}^{\mathrm{\,c}}\,\middle/\right.M_{\mathrm{1}}\right )^{\mathrm{5}}\, \right]\,\middle/\right.2.13,
  \end{equation}
where $M_{\rm 1}$ is the mass of the primordial primary, $M_{\rm 1}^{\,\rm c}$ is the core mass of the primordial primary, and $x={\rm d}\ln R_{\rm 1}/{\rm d}\ln M_{\rm 1}$ is the mass-radius exponent of the mass donor star and varies with its composition \citep{Hurley2002}.

For the treatment on the CE phase, we assume that a part of the orbital energy released by the system during the spiral-in process is injected into the envelope to eject the material in the CE \citep{Livio1988, Webbink1984, Han2004}. The CE is assumed to be ejected completely when $\alpha_{\rm CE}\,\Delta E_{\rm orb}=|E_{\rm bind}|$. Here, $E_{\rm bind}$ is the binding energy of the CE, $\Delta E_{\rm orb}$ is the orbital energy released during the spiral-in phase, and $\alpha_{\rm CE}$ is the CE ejection efficiency. In this work, we set $\alpha_{\rm CE}=1.0$ or $\alpha_{\rm CE}=3.0$ to investigate its influence on the results \citep{Meng2017}.

\subsubsection{Basic Assumptions}
\label{sec:assumptions}

We follow the evolution of $10^{\rm 7}$ binary systems which are generated in a Monte-Carlo way, starting with both stars at zero-age MS (ZAMS) until the eventual formation of He~star+MS system. The basic assumptions adopted in these rapid binary evolution calculations are given as follows:

\begin{itemize}
\item[(1)] A circular orbit is assumed for all binary systems. All stars are assumed to be members of binary systems\footnote{However, observational measurements of the binary fraction are still imprecise. Its realistic value is probably less than one. Therefore, the resultant SN Ia rate computed here is actually an upper limit.}.
\item[(2)] The primordial primary is generated according to the formula of \citet{Eggleton1989} which adopted a simple approximation to the initial mass function of \citet{Miller1979}:
\begin{equation}
M_{\rm 1}^{\rm p}=\frac{0.19X}{(1-X)^{\rm 0.75}+0.032(1-X)^{\rm
0.25}},
  \end{equation}
where $X$ is a random number in the range [0,1], and $M_{\rm1}^{\rm p}$ is the mass of the primordial primary, which is between 0.1\,$M_{\rm \odot}$ and 100\,$M_{\rm \odot}$.

\item[(3)] An uniform mass-ratio distribution is adopted, i.e., $q'$=$M_{\mathrm{2}}/M_{\mathrm{1}}$, is uniformly distributed in the range  $[0,1]$ \citep{Mazeh1992, Goldberg1994}:
\begin{equation}
n(q')=1, \hspace{2.cm} 0<q'\leq1.
\end{equation}

\item[(4)] For the separation  of the binary systems, a constant distribution in $\log\,a$ is assumed for wide binaries, while $a$ falls off smoothly for close binaries:
\begin{equation}
a\cdot n(a)=\left\{
 \begin{array}{lc}
 \alpha_{\rm sep}(a/a_{\rm 0})^{\rm m} & a\leq a_{\rm 0};\\
\alpha_{\rm sep}, & \ \ \ \ \ \ a_{\rm 0}<a<a_{\rm 1},\\
\end{array}\right.
\end{equation}
where $\alpha_{\rm sep}$=$0.07$, $a_{\rm 0}=10\,R_{\odot}$, $a_{\rm 1}=5.75\times 10^{\rm 6}\,R_{\odot}=0.13\,{\rm pc}$ and $m$=$1.2$. The separation distribution adopted here implies an equal number of wide binary systems per logarithmic interval, and gives approximately 50 percent of binary systems with an orbital period $\lesssim100$\,yr \citep{Han1995}.

\item[(5)] Either a single starburst, i.e., $10^{\rm11}\, M_{\odot}$ in stars is generated at a single instant of time, or a constant star-formation rate (SFR) of $5\,M_{\odot}\,{\rm yr^{-1}}$ for the last 15\,Gyr is adopted in this work \citep{Iben1984, Meng2017}. 

\end{itemize}

\begin{figure*}
  \begin{center}
    {\includegraphics[width=0.43\textwidth, angle=360]{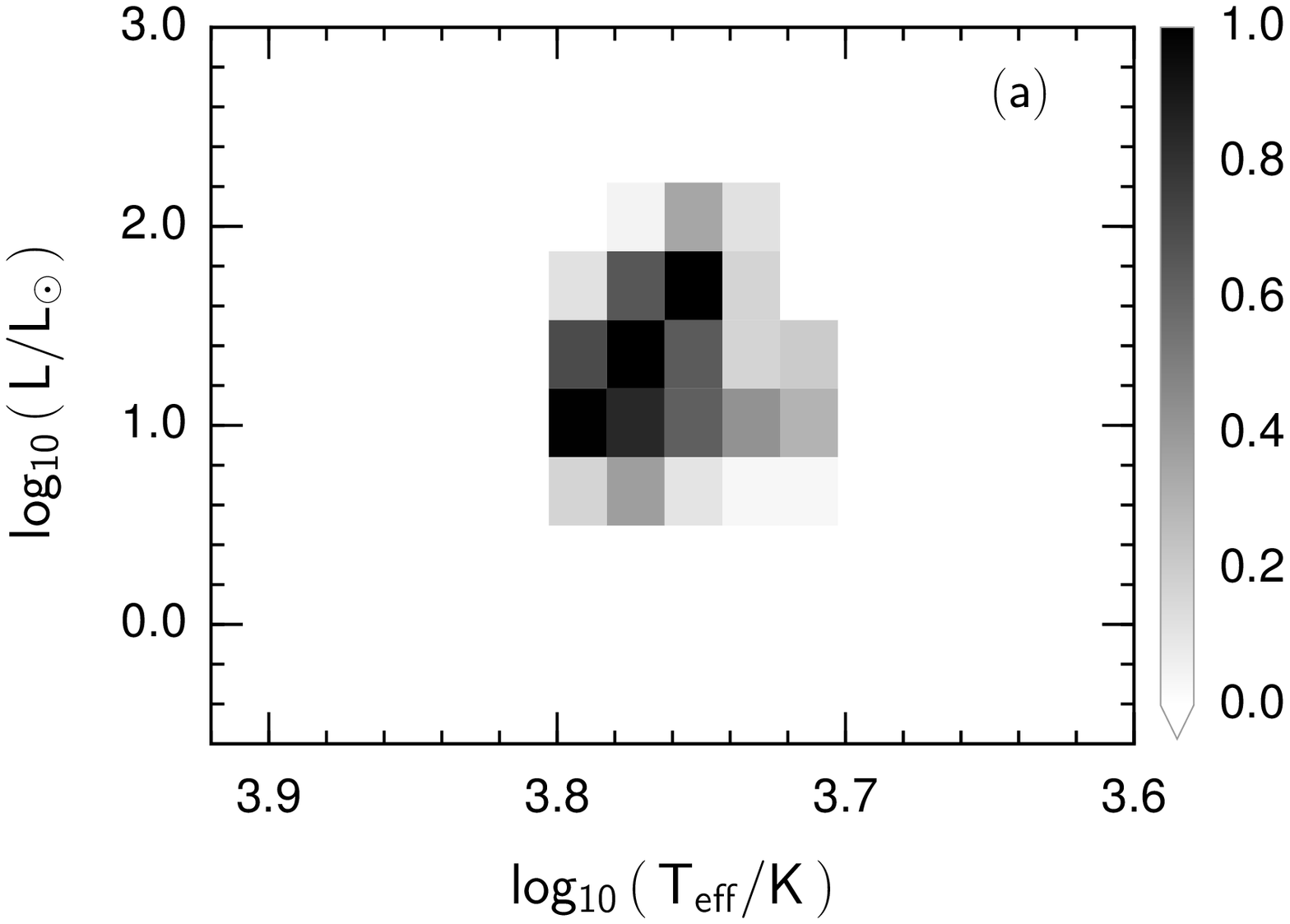}}
   \hspace{0.15in}
    {\includegraphics[width=0.43\textwidth, angle=360]{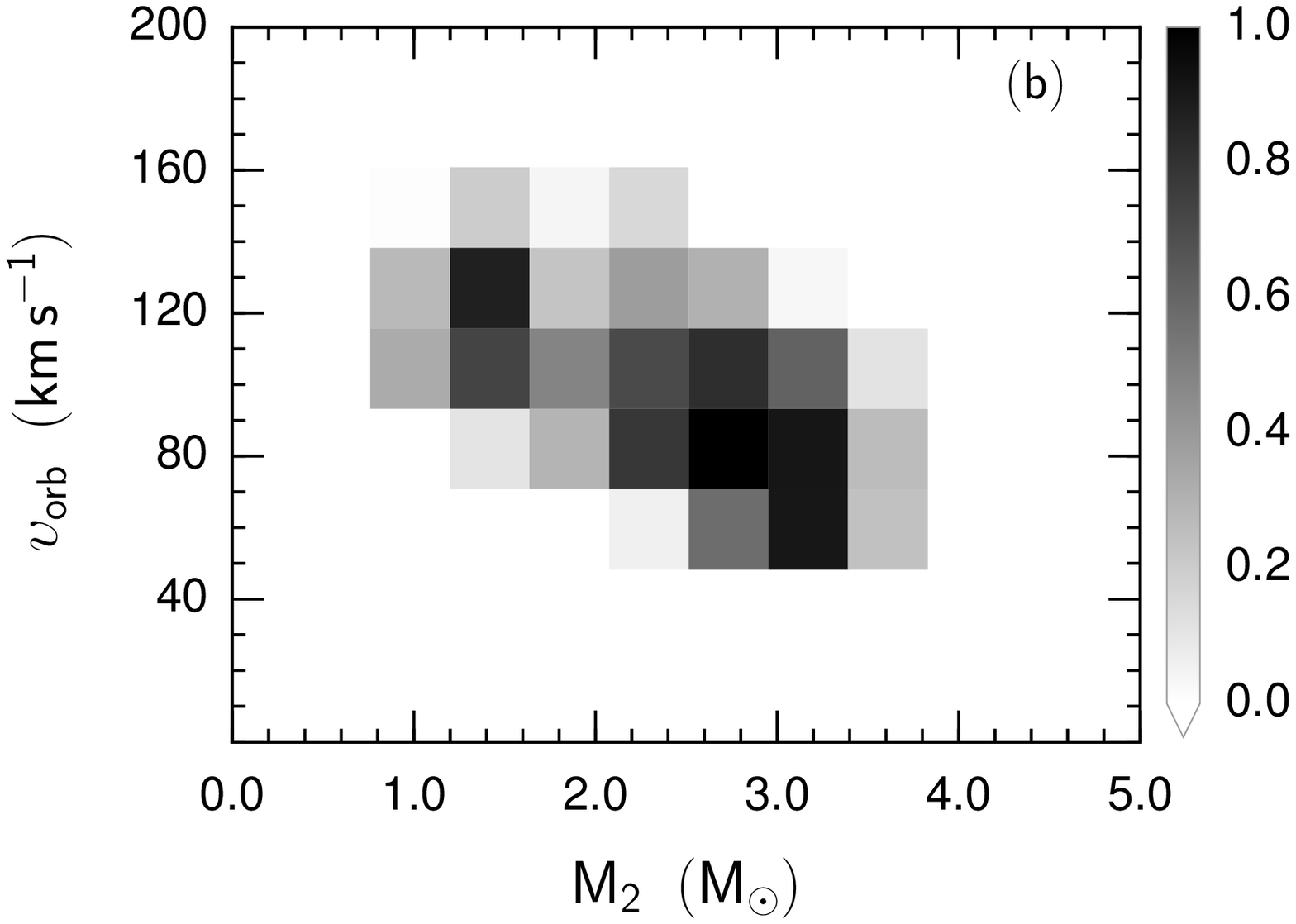}}
    {\includegraphics[width=0.43\textwidth, angle=360]{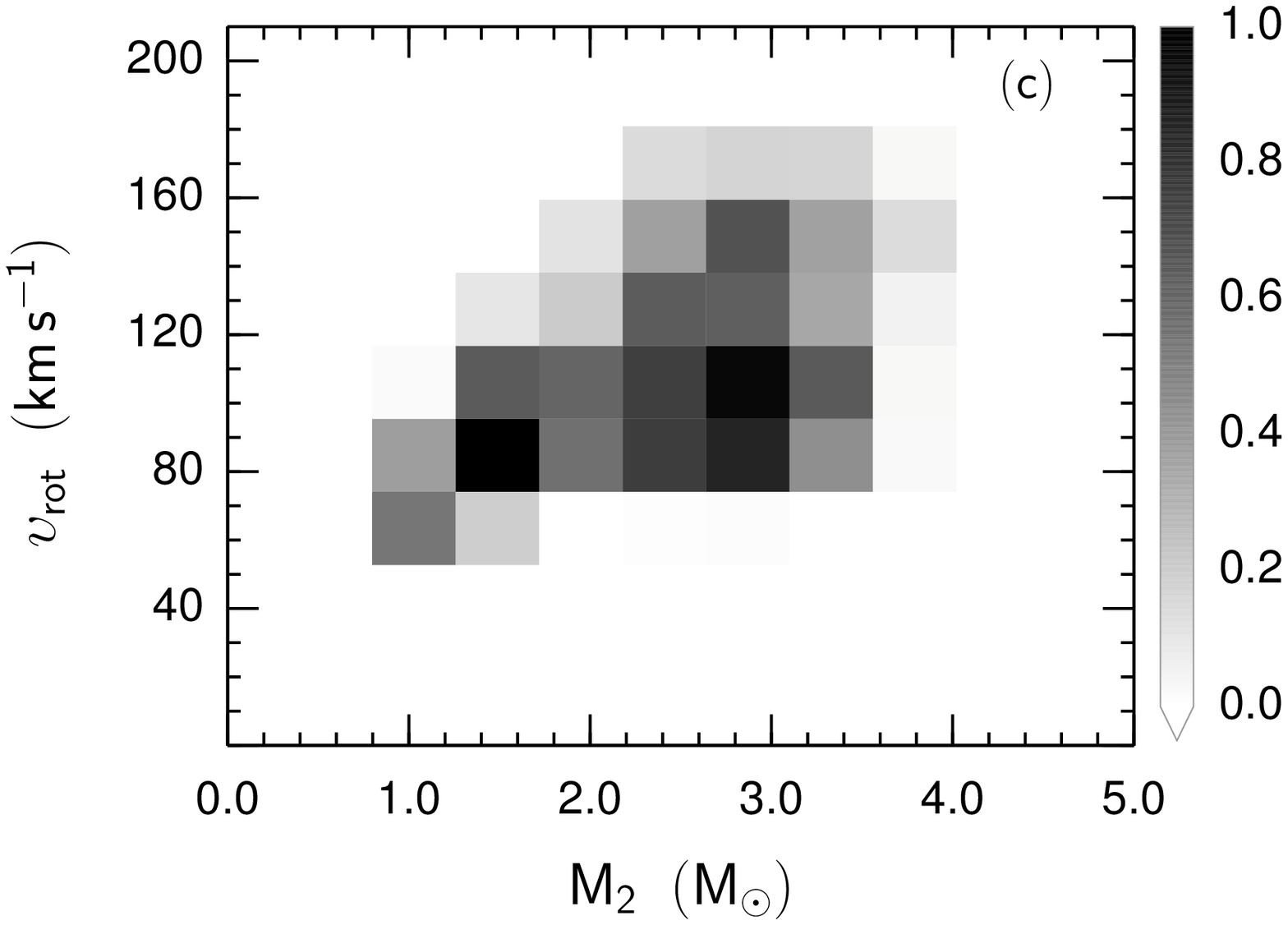}}
   \hspace{0.15in}
    {\includegraphics[width=0.43\textwidth, angle=360]{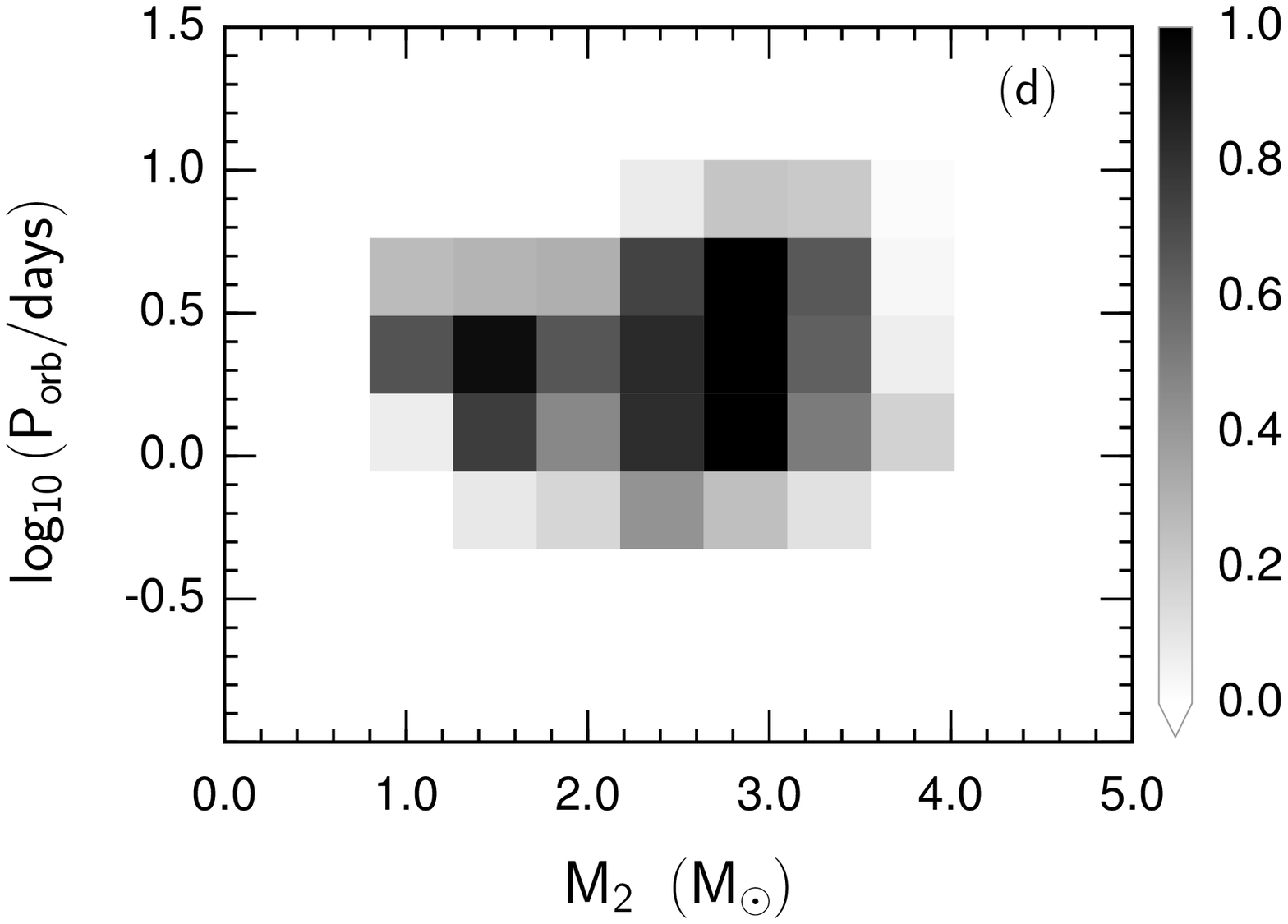}}
    \caption{Properties of the companion stars and binary orbital periods at the moment of SN Ia explosion in the HEMS donor scenario based on our population synthesis calculations with a CE ejection efficiency $\alpha_{\mathrm{CE}}$=$3.0$. Here, $L$, $T_{\mathrm{eff}}$, $M_{2}$, $\upsilon_{\mathrm{orb}}$ and $\upsilon_{\mathrm{rot}}$ are the luminosity, effective temperature, mass, orbital velocity and surface rotational velocity of the companion star, respectively. $P_{\mathrm{orb}}$ is the orbital period of the binary system.}
\label{Fig:pre-sne}
  \end{center}
\end{figure*}

\subsection{Rates and delay times of SNe Ia}
\label{sec:rates}

Fig.~\ref{Fig:rate} shows the evolution of rates of SNe Ia in the HEMS donor scenario for a constant SFR of $5.0\,M_{\sun}\,\rm{yr^{-1}}$ (left-hand panel) and for a single starburst of $10^{11}\,M_{\sun}$ (right-hand panel). As it shown, we find that the Galactic rate in this channel is around $0.6$--$1.2\times10^{-3}\,\rm{yr^{-1}}$ for different CE ejection efficiencies, which is lower than the observationally-inferred rate of $2.84\pm0.60\times10^{-3}\,\rm{yr^{-1}}$ \citep{Li2011a, Li2011b}. Our predicted rate is almost the same as that of the WD+MS channel in previous studies, $0.6$--$1.1\times10^{-3}\,\rm{yr^{-1}}$ \citep{Han2004}. This suggests that most companion stars in the WD+MS channel are HEMS rather than normal MS stars. In reality, the star-formation history of the Milky Way is different from a constant star-formation rate of $5.0\,M_{\sun}\,\rm{yr^{-1}}$ or a single starburst of $10^{11}\,M_{\sun}$, composed of different bulge and disk components \citep{Kubryk2015, Maoz2017}. For a comparison, we also adopt a more realistic SFR of the Milky Way given by \citet[][see their Fig.~2]{Kubryk2015}\footnote{Here, the SFR rises exponentially until $t\approx1.5\,\mathrm{Gyr}$ and peaks at a value of around $12\,M_{\sun}\,\mathrm{yr^{-1}}$, and then declines exponentially for about 10\,Gyr. The SFR is then a constant, giving the current SFR is about $2.0\,M_{\sun}\,\mathrm{yr^{-1}}$ \citep{Kubryk2015, Chomiuk2011}.} to predict the SN rates in the HEMS donor scenario (Fig.~\ref{Fig:rate}). With this more realistic SFR of the Milky Way, we obtain that the rates rise and reach a peak around $1.5\,\mathrm{Gyr}$, and then decline exponentially until $t\approx10\,\mathrm{Gyr}$. The SN rate is then approximately constant. As a result, the current SN rate in this case is lower than those with a constant SFR by a factor of 2.  

In addition, we find that the delay times of SNe Ia in the HEMS donor scenario cover a range from 0.1 to $1\,\mathrm{Gyr}$ after the burst (Fig.~\ref{Fig:rate}). A comparison between the observed delay-time distribution (DTD) of SNe Ia \citep{Maoz2010, Maoz2011, Maoz2012, Maoz2017, Graur2013} and our measured DTD is given in Fig.~\ref{Fig:rate}. It shows that the observed DTD covers a wide range which includes short delay times of a few $\times\,10\,\mathrm{Myr}$ and long delays of up to $10\,\mathrm{Gyr}$. The HEMS donor scenario cannot contribute SNe Ia which are younger than about $100\,\mathrm{Myr}$ and older than $1\,\mathrm{Gyr}$. It has been suggested that young (<$100\,\mathrm{Myr}$) and/or old (>$1\,\mathrm{Gyr}$) SNe Ia can be generated from CO WDs which accrete matter from a non-degenerate He star and/or a RG companion (e.g., \citealt{Ruit09, Ruit11, Wang2012, Claeys2014, Liu15a}). Also, population synthesis studies have presented that the predicted DTD in the DD scenario can almost cover the whole range of observed delays of SNe Ia \citep{Ruit09, Ruit11, Claeys2014}. 

Fig.~\ref{Fig:initial} presents the initial mass of a He star and mass ratio of the binary system at stage IV as a function of delay times in our population synthesis calculation for $\alpha_{\mathrm CE}$=$3.0$. The He~star+MS systems with a more massive initial He star tend to have a shorter delay time. Also, the mass ratios of He~star+MS systems at stage IV ($M_{\mathrm{MS}}$/$M_{\mathrm{He}}$) which successfully produce SNe Ia in the HEMS donor scenario are typically around $1.6$--$4.0$.

\section{Discussion}
\label{sec:discussion}

\subsection{Comparison with previous studies}
\label{sec:comparison}

A comparison between our results and those of the previous study by \citet{Han2004} is presented in Fig.~\ref{Fig:rate}. The distribution of delay times after the burst ($0.1$--$1\,\mathrm{Gyr}$) in our calculations is comparable with that of \citet{Han2004} who assumed CO~WD+MS binary systems have a normal solar-metallicity MS companion star. For the same CE ejection efficiency, $\alpha_{\mathrm CE}$=$1.0$, it is found that the HEMS donor scenario produces some younger SNe Ia than those of \citet{Han2004} because the HEMS stars generally evolve faster than normal MS stars to fill their roche-lobes. In addition, the Galactic rate of SNe Ia in the HEMS donor scenario for $\alpha_{\mathrm CE}$=$1.0$ is also found to be consistent to that of \citet{Han2004}.

In this work, we start our consistent binary evolution calculations with the He~star+MS binary system. However, \citet{Han2004} started their binary evolution calculations when CO~WD+MS binary systems are formed. In Fig.~\ref{Fig:binary}, we present the distributions of binary parameters of CO~WD+HEMS binary systems in our binary evolution calculations. These can compare with the distributions of initial parameters of binary evolution calculations by \citet{Han2004}, i.e., their initial contours for producing SNe Ia (see their Fig.~3). As shown, our CO~WD+HEMS binary systems which successfully produce SNe Ia cover a WD mass of $M_{\rm WD}\approx 0.8$--$1.1\,M_{\sun}$, a donor mass of $M_{\rm HEMS}\approx 1.5$--$4.6\,M_{\sun}$ and a orbital period of $\mathrm{log_{10}}\,(P_{\rm orb}\,\mathrm{d^{-1})}\approx-0.3$--$1.4$. In \citet{Han2004}, their initial contours cover a WD mass of $M_{\rm WD}^{\rm i}\approx 0.67$--$1.2\,M_{\sun}$, a donor mass of $M_{\rm 2}^{\rm i}\approx 1.8$--$3.7\,M_{\sun}$ and an orbital period of $\mathrm{log_{10}}\,(P^{\rm i}\,\mathrm{d^{-1})}\approx-0.3$--$1.2$ (see their Fig.~3). In our HEMS donor scenario, we cannot produce SNe Ia with a initial WD star less massive than $0.8\,M_{\sun}$ or more massive than $1.1\,M_{\sun}$, which is strongly constrained by the mass of a He star in He~star+MS systems at the initial phase of our calculations, i.e., $M_{1}^{\rm i}$=$0.85$--$1.45\,M_{\sun}$ (Figs.~\ref{Fig:1} and~\ref{Fig:contour}).

As mentioned above, there are three channels, i.e., the $HG/RGB$, $EAGB$ and $TPAGB$ channel, are proposed for forming CO~WD+HEMS/MS binary systems (Section~\ref{sec:introduction}). In this work, we only include the $HG/RGB$ channel as it is the most efficient channel for producing SNe Ia among these three evolutionary channels \citep{Wang2012}. However, the $TPAGB$ channel was also included into binary population synthesis calculations of \citet{Han2004}. If the $EAGB$ and $TPAGB$ channel are included into our calculations, we expect that the Galactic rate of SNe Ia would be slightly higher than that in Fig.~\ref{Fig:rate}. It has been suggested that the contribution of these two channels to the total SNe Ia in the CO~WD+MS scenario is $\lesssim\,45\%$ \citep{Wang2012}.

\begin{figure}
  \begin{center}
    {\includegraphics[width=0.45\textwidth, angle=360]{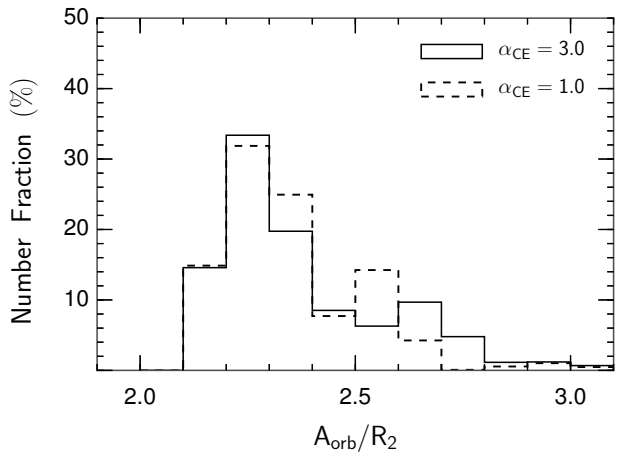}}
    \caption{Distributions of the ratio of binary separation to companion radius ($A_{\mathrm{orb}}/R_{2}$) at the moment of SN Ia explosion based on our population-synthesis calculations.}
\label{Fig:ar}
  \end{center}
\end{figure}

\subsection{Pre-explosion companion stars}
\label{sec:pre-SN}

In the SD scenario, the pre-explosion signatures of progenitor systems are generally determined by the non-degenerate companion stars because the WDs are faint and they can only be observed directly in our own Milky Way and very nearby galaxies. Therefore, analyzing pre-explosion images at the SN position provides a direct way to constrain the nature of the SN Ia companion star \citep{Li2011}. However, no progenitors of normal SNe Ia have yet been directly observed to date, even for the detection of the relatively nearby normal SNe Ia, SN~2011fe and SN~2014J (e.g., \citealt{Li2011, Bloom2012, McCully2014, Kell14}). However, the probable companion star of the progenitor system of a Iax event (see Section~\ref{sec:Iax}) SN~2012Z, i.e., SN~2012Z-S1: $\rm{log_{10}}\,(T_{\rm{eff}}/\rm{K})\approx4.07$, $\rm{log_{10}}\,(L/L_{\sun})\approx4.23$, has been recently discovered from pre-explosion $\textit{Hubble Space Telescope}$ images \citep{McCully2014}. It is further suggested that SN~2012Z had a progenitor system consisting of a He~star donor and a WD \citep{McCully2014, Liu15b}.

Providing theoretical predictions for observable properties of pre-explosion companion stars is needed for identifying them in pre-explosion images  \citep{Han2008}. Fig.~\ref{Fig:pre-sne} gives the properties (the luminosity $L$, effective temperature $T_{\mathrm{eff}}$, mass $M_{2}$, orbital velocity $\upsilon_{\mathrm{orb}}$ and surface rotational velocity $\upsilon_{\mathrm{rot}}$) of the companion stars, and the orbital periods of the systems ($P_{\mathrm{orb}}$) at the moment of SN explosion from our populations synthesis calculations for $\alpha_{\mathrm CE}$=$3.0$. Although the companion properties in the HEMS donor scenario cannot explain the pre-explosion observations of SN~2012Z, comparing our results with future SN Ia pre-explosion observations is expected to help with identifying the companion stars of SN Ia progenitors and thus examining the validity of the HEMS donor scenario.

\subsection{Surviving companion stars}
\label{sec:survivor}

In the SD scenario, as the SN explodes, the companion star is released from its orbit and ejected with a moving velocity mainly dominated by its pre-explosion orbital velocity. Searching for the companion star in SN Ia remnants provides a promising way to identify the SD progenitor system.

Fig.~\ref{Fig:pre-sne} shows that the companion stars in the HEMS donor scenario have an orbital velocity,  $\upsilon_{\mathrm{orb}}\approx50$--$160\,\mathrm{km\,s^{-1}}$ and a surface rotational velocity of $\upsilon_{\mathrm{rot}}\approx50$--$180\,\mathrm{km\,s^{-1}}$ at the moment of the SN explosion\footnote{Here, we assume that the donor star has a spin corresponding to the orbital frequency of the binary system because the rotation of the donor star is expected to be tidally locked to its orbital motion.}. While interacting with the SN ejecta, the companion star receives a kick velocity which is typically about $\upsilon_{\mathrm{kick}}\gtrsim$ a few\,$\times\,50\,\mathrm{km\,s^{-1}}$ \citep{Liu2012}. One therefore can roughly predict that the surviving companion star in the HEMS donor scenario will have a spatial velocity of $\upsilon_{\mathrm{spatial}}$=$\sqrt{\upsilon_{\mathrm{orb}}^{2}+\upsilon_{\mathrm{kick}}^{2}}\gtrsim 50$--$170\,\mathrm{km\,s^{-1}}$. Meanwhile,  after the SN impact, the companion star is significantly shocked and heated while a part of its outer layers is removed. As a result, the companion star significantly puffs up and the stellar envelope is out of thermal equilibrium \citep{Liu2012}. It has been suggested that the equatorial surface rotational velocity of the companion star can dramatically drop to a few $\times\,10\%$ of the original value during its re-equilibration phase on the Kelvin-Helmholtz timescale of around $10^{3}$--$10^{5}$ yrs \citep{Liu2013a}. Also, with tracing the long-term evolution of the companion star, it has been suggested that the star should be overluminous during its re-equilibration phase \citep{Pan13}. But unfortunately, to date, no surviving companion stars have been found in SN Ia remnants (e.g., \citealt{Ruiz04, Ruiz2017, Kerz09, Kerz13, Kerz14, Kerzendorf2017a, Kerzendorf2017b, Scha12, Bedin2014}).

\subsection{Detection of stripped hydrogen}
\label{sec:hydrogen}

The SN ejecta interacts with the companion star, removing material from its surface. If the removed companion material contains a large amount of H, some signatures of H would be shown in SN Ia nebular spectra \citep{Leon07}. By combining the ratio of binary separation to companion radius ($A_{\mathrm{orb}}/R_{2}$) at the moment of SN Ia explosion obtained from consistent binary evolution calculations with a hydrodynamical model, \citet{Liu2017} predicted that the swept-up hydrogen masses expected in the HEMS donor scenario are around $0.10$--$0.17\,M_{\sun}$.

Fig.~\ref{Fig:ar} shows that the distribution of $A_{\mathrm{orb}}/R_{2}$ from our population synthesis calculations peaks around 2.3. As done in \citet{Liu2017}, using the power-law relationship between stripped companion mass and $A_{\mathrm{orb}}/R_{2}$ (see their Eq.~2) derived from hydrodynamical simulations by \citet{Liu2012}, we can estimate that the amount of stripped H in the HEMS donor scenario is typically around $0.15\,M_{\sun}$, which is much higher than observational upper-limits on the H masses in progenitor systems of SNe Ia, i.e., $0.001$--$0.058\,M_{\sun}$ \citep{Maguire2016}. This suggests that most of SNe Ia generated from the HEMS donor scenario would likely present some features of stripped H in their late-time (nebular) spectra.

\subsection{Progenitors of SNe Iax?}
\label{sec:Iax}

SNe Iax have been proposed as one new sub-class of SNe Ia \citep{Fole13, Jha2017}. They are significantly fainter than normal SNe Ia. They have a wide range of explosion energies ($10^{49}$--$10^{51}\,\rm{erg}$), ejecta masses ($0.15$--$0.5\,M_{\sun}$), and $^{56}\rm{Ni}$ masses ($0.003$--$0.3\,M_{\sun}$). Comparing to normal SNe Ia ($15000\,\rm{km\,s^{-1}}$), the spectra of SNe Iax are characterized by lower expansion velocities ($2000$--$8000\,\rm{km\,s^{-1}}$) at similar epochs. Strong He lines were identified in spectra of two SNe Iax, i.e., SN~2004cs and SN~2007J \citep{Fole09, Fole13}. The rate of occurrence of SNe Iax is estimated to be about $5$--$30\%$ of the total SN Ia rate \citep{Fole13}. The progenitor systems of SNe Iax are suggested to have relatively short delay times ($<500\,\rm{Myr}$) because most SNe Iax are observed in late-type, star-forming galaxies \citep{Fole13, Lyma13, Whit14}. However, there is at least one Iax event, SN 2008ge, which was observed in an S0 galaxy with no signs of star formation, indicating a long delay time \citep{Fole13}.

Some recent studies have suggested that SNe Iax are more likely to be produced from the weak pure deflagration explosion models of Chandrasekhar-mass WDs in the SD progenitor scenario \citep{McCully2014, Kromer2015, Kromer2016, Vennes2017}. As shown in Section~\ref{sec:rates}, the Galactic rate of SNe Ia in the HEMS donor scenario of $0.6$--$1.2\times10^{-3}\,\rm{yr^{-1}}$ is comparable to the estimated rate of SNe Iax. However, the delay-time distributions in this scenario (0.1--$1\,\mathrm{Gyr}$) struggle to explain the fact that most SNe Iax have a short delay time. Also, as shown in Section~\ref{sec:pre-SN}, pre-explosion companion properties predicted from the HEMS donor scenario cannot provide an explanation for the blue source detected in pre-explosion observations of SN~2012Z \citep{McCully2014}. We therefore conclude that the HEMS donor scenario is unlikely to be a common progenitor scenario for SNe Iax.

\subsection{Imprint of accretion winds}
\label{sec:wind}

Our HEMS donor models adopt the method of \citet{Hachisu1999} to calculate the mass growth rate of the WD. The so-called optically thick ``accretion wind'' outflows is assumed to be driven from the WD surface if the mass transfer rate is higher than the critical accretion rate for stable H burning (Section~\ref{sec:binary}). These optically thick outflows from the WD surface modify the structure of the CSM at the time of the SN explosion, evacuating a low-density detectable cavity around the WD \citep{Badenes2007}. Also, in the SD scenario the accreting WDs are a powerful source to photoionize \Heii in the surrounding gas, producing emission in the \Heii recombination line at $\lambda4686\,\AA$. These \Heii recombination lines can be used as a test of the nature of SN Ia progenitors \citep{Woods2013, Johansson2014, Graur2014b}. Unfortunately, to date, neither of a cavity around the WD nor \Heii recombination lines have been detected \citep{Badenes2007, Johansson2014, Graur2014}. This seems to present a challenge to the SD donor scenario of SNe Ia. However, more observations and improvements of proposed SN Ia progenitor models are still required to give a stronger conclusion.

\subsection{Model uncertainties}

In our detailed binary evolution, we use the prescription of the optically thick wind model of \citet{Hachisu1999} and He-retention efficiencies from \citet{Kato1999} to describe the mass accumulation efficiency of accreting WDs. However, the exact mass-retention efficiency is still not well-constrained. It has been suggested that the results might be different from those in our binary population synthesis calculations if different mass-retention efficiencies are adopted (e.g., \citealt{Pier14, Ruit14, Toonen2014}). 

It should be kept in mind that the initial conditions and assumptions of binary population synthesis calculations such as current star-formation rate, initial mass function and the CE evolution are still weakly constrained. This leads to some uncertainties on the results of population synthesis calculations \citep{Toonen2014, Claeys2014}. We refer the reader to \citet{Claeys2014} for a detailed discussion for the effect of different theoretical uncertainties in population synthesis models of SNe Ia.

\section{Summary}
\label{sec:summary}

In this work, we have investigated the rates and delay times of SNe Ia in the HEMS donor scenario by combining the results of consistent binary evolution calculations into population synthesis models. Our main results and conclusions can be summarized as follows: 

\begin{itemize}\itemsep5pt 
\item  The theoretical Galactic rate of SNe Ia in the HEMS donor scenario is about $0.6$--$1.2\times10^{-3}\,\rm{yr^{-1}}$, which is about 30\% of that inferred observationally. Also, delay times of SNe Ia in this scenario cover a wide range from 0.1 to $1\,\mathrm{Gyr}$. Both the Galactic rates and delay times are consistent with those of the previous study for the CO~WD+MS channel \citep{Han2004}.

\item  Companion properties at the moment of SN explosion in the HEMS donor scenario are given in Fig.~\ref{Fig:pre-sne}, which will be helpful for identifying the companion stars of SN Ia progenitor systems in their pre-explosion images.

\item  Our population-synthesis calculations predict that the amount of H mass removed by SN explosion from the companion star in the HEMS donor is typically massive than $0.1\,M_{\sun}$.

\item  Although the SN rates in the HEMS donor scenario are comparable to those of SNe Iax, delay times of SNe and pre-explosion companion properties in this scenario are difficult to provide an explanation for current observations of SNe Iax. We therefore conclude that the HEMS donor scenario is unlikely to be a common progenitor scenario for SNe Iax.

\end{itemize}

\section*{Acknowledgements}

We thank the anonymous referee for his/her valuable comments that helped to improve the paper. We would like to thank Carlo Abate for his fruitful discussions. Z.-W.L thanks Bo Wang and Hai-Liang Chen for their suggestions on population synthesis calculations. This work is supported by the Alexander von Humboldt Foundation. R.J.S. is the recipient of a Sofja Kovalevskaja Award from the Alexander von Humboldt Foundation.

\bibliographystyle{mnras}

\bibliography{ref}

\end{document}